\pdfoutput=1
\documentclass[conference]{IEEEtran}

\usepackage{etex}
\usepackage{amssymb}
\usepackage{times}
\usepackage{bm}
\usepackage{subcaption}
\usepackage{amsmath}
\usepackage{graphicx}
\usepackage{amssymb}
\usepackage{amstext}
\usepackage{latexsym}
\usepackage{color,soul}
\usepackage{ifthen}
\usepackage{multirow}
\usepackage{verbatim}
\usepackage{array,tabularx}
\usepackage{accents}
 \usepackage{cite}
 \usepackage{arydshln}
\usepackage{hyperref}
\usepackage{url,amssymb,epsfig,times,mathrsfs,algorithm,algorithmic,latexsym,xspace,fancyhdr,wrapfig, xcolor,multirow,amsthm}
\usepackage{caption}
\usepackage{soul}
\usepackage[inline]{enumitem}
\usepackage{arydshln}
\usepackage{enumitem}
\usepackage[makeroom]{cancel}
\usepackage{mathtools}
\usepackage{dblfloatfix}
\usepackage[utf8]{inputenc}

\usepackage{tikz}
\usetikzlibrary{shapes.geometric}
\usetikzlibrary{shapes,snakes,patterns}
\usetikzlibrary{arrows,positioning,automata,calc}
\usetikzlibrary{plotmarks}
\tikzset{
    int/.style={
           rectangle,
           rounded corners,
           draw=black, thin, fill=black!20,
           minimum height=2em,
           inner sep=2pt,
           text centered,
           },
}

\newtheorem{theorem}{Theorem}

\newtheorem{claim}{Claim}
\newtheorem{example}{Example}
\newtheorem{definition}{Definition}
\newtheorem{remark}{Remark}

\IEEEoverridecommandlockouts

\pdfoutput=1

\begin{document}
\pdfoutput=1
\allowdisplaybreaks
\newlength\figureheight
\newlength\figurewidth

\title{ Guess \& Check Codes for Deletions and Synchronization \vspace{-0.2cm} \thanks{The authors are with the ECE department of Illinois Institute of Technology, Chicago. Emails: skashann@hawk.iit.edu and salim@iit.edu.}
\thanks{This work was supported  in parts by NSF Grant CCF 15-26875.} }
\author{
\IEEEauthorblockN{Serge Kas Hanna,  Salim El Rouayheb 
}
\vspace{-0.5cm}
}
\maketitle
\begin{abstract} We consider the problem of constructing codes that can correct $\delta$ deletions occurring in an arbitrary binary string of length $n$ bits. Varshamov-Tenengolts (VT) codes can correct all possible single deletions $(\delta=1)$ with an asymptotically optimal redundancy. Finding similar codes for $\delta \geq 2$ deletions is an open problem. We propose a new family of codes, that we call Guess \& Check (GC) codes, that can correct, with high probability, a constant number of deletions $\delta$ occurring at uniformly random positions within an arbitrary string. The GC codes are based on MDS codes and have an asymptotically optimal redundancy that is $\Theta(\delta \log n)$. We provide deterministic polynomial time encoding and decoding schemes for these codes. We also describe the applications of GC codes to file synchronization. 
\end{abstract}

%
%
%
%

\section{Introduction}
The  deletion channel is probably the most notorious example of a point-to-point channel whose capacity remains unknown. The bits that are deleted by this channel are completely removed from the transmitted sequence and their locations are unknown at the receiver (unlike erasures). For example, if $1010$ is transmitted, the receiver would get $00$ if the first and third bits were deleted. Constructing efficient codes for the deletion channel  has also been a challenging task. Varshamov-Tenengolts (VT) codes \cite{VT65} are the only deletion codes with asymptotically optimal redundancy and can correct only a single deletion. The study of the deletion channel has many applications such as file synchronization \cite{V15,Y14} and DNA-based storage \cite{RO}. 





The capacity of the deletion channel has been studied in the probabilistic model where the deletions are i.i.d. and occur with a fixed probability $p$. An immediate upper bound on the channel capacity is given by the capacity of the erasure channel $1-p$.  Mitzenmacher and Drinea showed in \cite{M06} that the capacity is at least $(1-p)/9$. Extensive work in the literature has focused on determining lower and upper bounds on the capacity \cite{M06,D07,Ven13,K13}. We refer interested readers to the  comprehensive survey by Mitzenmacher \cite{M09}.

A separate line of work has focused on constructing codes that can correct a given number of deletions. In this work we are interested in binary codes that correct a constant number of deletions $\delta$. Levenshtein  showed in \cite{L66} that VT codes \cite{VT65} are capable of correcting a single deletion ($\delta=1$), with an asymptotically optimal redundancy ($\log(n+1)$ bits). 
VT codes have been used to construct codes that can correct a combination of a single deletion and multiple adjacent tranpositions \cite{RO}. However, finding VT-like codes for multiple deletions ($\delta \geq 2$) is an open problem.
In \cite{L66}, Levenshtein provided bounds showing that the asymptotic number of redundant bits needed to correct $\delta$ bit deletions in an $n$ bit codeword is $\Theta(\delta \log n)$, i.e., $c~\delta \log n$ for some constant $c>0$. Levenshtein's bounds were later generalized and improved in \cite{N14}.

The simplest code for correcting $\delta$ deletions is the $(\delta+1)$ repetition code, where every bit is repeated $(\delta+1)$ times. However, this code is inefficient because it requires $\delta n$ redundant bits, i.e., a redundancy that is linear in $n$. Helberg codes \cite{H02} are a generalization of VT codes for multiple deletions. These codes can correct mutiple deletions but their redundancy is at least linear in $n$ even for two deletions. Schulman and Zuckerman in \cite{S99} presented codes that can correct a constant fraction of deletions. Their construction was improved in \cite{G14}, but the redundancies in these constructions are $\mathcal{O}(n)$.
Recently in \cite{B16}, Brakensiek {\em et al.} provide an explicit encoding and decoding scheme, for fixed $\delta$, that has $\mathcal{O}(\delta^2 \log \delta \log n)$ redundancy and a near-linear complexity. But the crux of the approach in \cite{B16} is that the scheme is limited to a specific family of strings, which the authors in \cite{B16} refer to as {\em pattern rich} strings. 
In summary, even for the case of two deletions, there are no known explicit codes for arbitrary strings, with $\mathcal{O}(\delta \log n)$ redundancy. 
\begin{figure*}
\centering
\resizebox{1\textwidth}{!}{
\begin{tikzpicture}[node distance=2.5cm,auto,>=latex']
\draw (-0.6,0) -- (-0.45,0);
    \node [int] (c) [text width=0.8cm,align=center]{\scriptsize Binary to $q-$ary};
    \node (b) [left of=c,node distance=1.5cm, coordinate] {a};
    \node [int] (z) [right of=c, node distance=3.4cm,text width=2.75cm,align=center] {\scriptsize Systematic MDS $\left(k/\log k+c,k/\log k \right)$};
    \node [int] (y) [right of=z, node distance=3.8cm,text width=0.8cm,align=center] {\scriptsize $q-$ary to binary};
    \node [int] (y1) [right of=y, node distance=3cm, text width=1.5cm,align=center] {\scriptsize {$(\delta+1)$ repetition of parity bits}};
    \node [coordinate] (end) [right of=c, node distance=2cm]{};
    \path[->] (b) edge node {\scriptsize $\mathbf{u}$} node[below] {\scriptsize $k$ bits} (-0.6,0);
    
    \draw[->] (c) edge node {\scriptsize $\mathbf{U}$} (z) ;
     \node(h1) [right of=c,node distance=1.2cm] [below] {\scriptsize $k/\log k$};
    \node(h2) [below of=h1,node distance=0.3cm] {\scriptsize symbols};
    \node(q1) [below of=c,node distance=0.8cm] {\scriptsize $q=k$};
    \node(q2) [below of=y,node distance=0.8cm] {\scriptsize $q=k$};
    \path[->] (z) edge node {\scriptsize $\mathbf{X}$} (y); 
    \node(i1) [right of=z,node distance=2.37cm] [below] {\scriptsize $k/\log k+c$};
   \node(name)[above of=i1,node distance=1.2cm] [above] {\scriptsize Guess \& Check (GC) codes};
    \node(i2) [below of=i1,node distance=0.3cm] {\scriptsize symbols};
    \node(e1) [right of=y1,node distance=3.3cm] {};
    \path[->] (y1) edge node {\scriptsize $\mathbf{x}$} (e1);
    \path[->] (y) edge node {} (y1);
    \node(t1) [right of=y,node distance=1.33cm] [below] {\scriptsize $k+c\log k$};
    \node(t2) [below of=t1,node distance=0.3cm] {\scriptsize bits};
    \node(tt1) [right of=y1,node distance=2.05cm] [below] {\scriptsize $k+c(\delta +1)\log k$};
    \node(tt2) [below of=tt1,node distance=0.3cm] {\scriptsize bits}; 
    \node(b1) [above of=c,node distance=0.8cm] {\scriptsize Block I};
    \node(b2) [above of=z,node distance=0.565cm] {\scriptsize Block II};
    \node(b3) [above of=y,node distance=0.8cm] {\scriptsize Block III};
    \node(b4) [above of=y1,node distance=0.8cm] {\scriptsize Block IV};
    
    \draw[dashed] (-0.6,-1) rectangle (11.1,1);
    
\end{tikzpicture} }
\captionsetup{font=footnotesize}
\caption{General encoding block diagram of the GC code for $\delta$ deletions. Block I: The binary message of length $k$ bits is chunked into adjacent blocks of length $\log k$ bits each, and each block is mapped to its corresponding symbol in $GF(q)$ where $q=2^{\log k}=k$. Block II: The resulting string is coded using a systematic $\left(k/\log k+c,k/\log k \right)$ $q-$ary MDS code where $c>\delta$ is the number of parity symbols. Block III: The symbols in $GF(q)$ are mapped to their binary representations. Block IV: Only the parity bits are coded using a $(\delta+1)$ repetition code.}
\label{fig:1}
\vspace{-0.5cm}
\end{figure*}
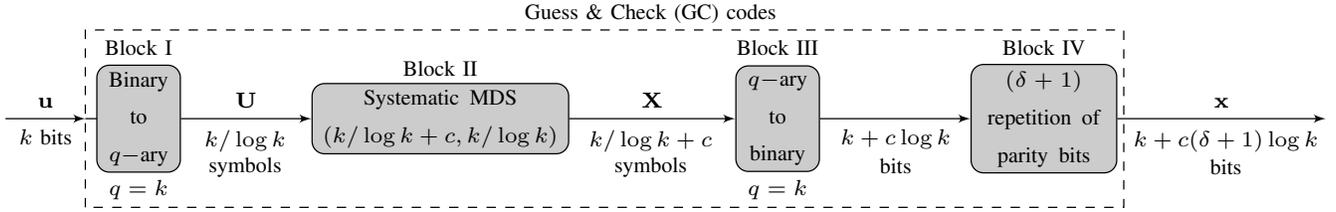

{\em Contributions:} While the work on codes that correct multiple deletions in \cite{H02,S99,G14,B16} focuses on zero-error codes, in our approach we relax this requirement and allow an asymptotically vanishing probability of decoding failure\footnote{The term {\em decoding failure} means that the decoder cannot make a correct decision and outputs a ``failure to decode" error message.}. Our contributions are the following: (i)~we propose new explicit codes, which we call  Guess \& Check (GC) codes,  that can correct, with high probability, and in polynomial time, a constant number of deletions\footnote{The GC code construction can be modified to correct $\delta$ insertions (instead of $\delta$ deletions). However, we focus here only on deletions.} $\delta$ occurring at uniformly random positions within an arbitrary binary string. The GC codes have an asymptotically optimal redundancy of value $c(\delta+1)\log k \approx c(\delta+1)\log n$ (asymptotically), where $k$ and $n$ are the lengths of the message and codeword, respectively, and $c>\delta$ is a constant integer; (ii)~GC codes enable different trade-offs between redundancy, decoding complexity, and probability of decoding failure; (iii)~we provide numerical simulations on the decoding failure of GC codes. Moreover, we describe how to use them for file synchronization as part of the interactive algorithm proposed by Venkataramanan {\em et al.} in \cite{V15} and provide simulation results highlighting the resulting savings in number of rounds and total communication cost.

\vspace{-0.2cm}
\section{Notation}
\label{sec:2}
Let $k$ and $n$ be the lengths in bits of the message and codeword, respectively.  Let $\delta$ be the number of deletions. We assume WLOG that $k$ is a power of $2$. Our code is based on a $q-$ary systematic $\left(\left \lceil k/\log k \right \rceil + c,\left \lceil k/\log k \right \rceil \right)$ MDS code, where $q=k>\left \lceil k/\log k \right \rceil + c$ and $c>\delta$ is a code parameter representing the number of MDS parity symbols.
We will drop the ceiling notation for $\left \lceil k/\log k \right \rceil$ and simply write $k/\log k$. All logarithms in this paper are of base $2$. The block diagram of the encoder is shown in Fig.~\ref{fig:1}. We denote binary and $q-$ary  vectors by lower and upper case bold letters respectively, and random variables by calligraphic letters.

\section{Main Result}
\label{sec:3}
Let $\mathbf{u}$ be a binary vector of length $k$ with i.i.d. Bernoulli$(1/2)$ components representing the information message. The message $\mathbf{u}$ is encoded into the codeword $\mathbf{x}$ of length $n$ bits using the Guess \& Check (GC) code illustrated in Fig.~\ref{fig:1}. 

\begin{theorem} 
\label{thm:1} 
The Guess \& Check (GC) code can correct in polynomial time up to a constant number of $\delta$ deletions occurring at uniformly random positions within~$\mathbf{x}$. Let $c>\delta$ be a constant integer. The code has the following properties:
\begin{enumerate}[leftmargin=*]
\item Redundancy: $n-k = c(\delta+1) \log k$ bits. 
\item Encoding complexity is $\mathcal{O}(k \log k)$, and  decoding complexity is $\mathcal{O}\left(\frac{k^{\delta+2}}{\log^{\delta}k}\right)$.
\item Probability of decoding failure: $Pr(F) = \mathcal{O} \left( \frac{k^{2\delta-c}}{\log^{\delta} k} \right)$. 
\end{enumerate}
\end{theorem}

\noindent GC codes enable trade-offs between the properties above, this will be highlighted later in Remark~\ref{rem1}. These properties show that: (i)~the code rate, $R=k/(k+c(\delta+1) \log k)$, is asymptotically optimal and approaches one as $k$ goes to infinity; (ii)~the order of complexity is polynomial in $k$ and is not affected by the constant $c$; (iii)~the probability of decoding failure goes to zero polynomially in $k$ if $c>2\delta$; and exponentially in $c$ for a fixed $k$.
Note that the decoder can always detect when it cannot decode successfully. This can serve as an advantage in models which allow feedback. There, the decoder can ask for additional redundancy to be able to decode successfully.  
\vspace{-0.15cm}
\section{Examples} \label{sec:4} The GC code we propose can correct  up to $\delta$ deletions with high probability. We provide examples  to illustrate the encoding and decoding schemes. The examples are for $\delta=1$ deletion just for the sake of simplicity\footnote{VT codes can correct one deletion with zero-error. However, GC codes are generalizable to multiple deletions.}. 
\begin{example}[Encoding]
\label{ex:1} Consider a message $\mathbf{u}$  of length $k=16$ given by $\mathbf{u}=1110000011010001$.
$\mathbf{u}$ is encoded by following the different encoding blocks illustrated in Fig.~\ref{fig:1}. \\
\noindent $1)$ {\em Binary to $q-$ary (Block I, Fig.~\ref{fig:1})}. The message $\mathbf{u}$ is chunked into adjacent blocks of length $\log k=4$ bits each, 
\begin{equation*}
\mathbf{u}=\underbrace{\overbracket{1~1~1~0}^{\text{block 1}}}_{\alpha^{11}}~\underbrace{\overbracket{0~0~0~0}^{\text{block 2}}}_{0}~\underbrace{\overbracket{1~1~0~1}^{\text{block 3}}}_{\alpha^{13}}~\underbrace{\overbracket{0~0~0~1}^{\text{block 4}}}_{1}\color{black}.
\end{equation*}
Each block is then mapped to its corresponding symbol in $GF(q)$, $q=k=2^4=16$. This results in a string $\mathbf{U}$ which consists of $k/\log k=4$ symbols in $GF(16)$. The extension field used here has a primitive element $\alpha$, with $\alpha^4=\alpha+1$.  Hence, 
we obtain $\mathbf{U}=(\alpha^{11},0,\alpha^{13},1\color{black})\in GF(16)^4$. \\
$2)$ {\em Systematic MDS code (Block II, Fig.~\ref{fig:1})}. $\mathbf{U}$ is then coded using a systematic $(k/\log k +c,k/\log k)=(6,4)$ MDS code over $GF(16)$, with $c=2>\delta$. The first parity is the sum of the systematic symbols and the encoding vector for the second parity is chosen to be $\left(1,\alpha,\alpha^{2},\alpha^{3}\right)$. The encoded string is $\mathbf{X}= (\color{blue} \alpha^{11},0,\alpha^{13},1,\color{red} \alpha,\alpha^{10} \color{black})$.\\
$3)$ {\em $Q-$ary to binary (Block III, Fig.~\ref{fig:1})}. The binary codeword corresponding to $\mathbf{X}$, of length $n=k+2\log k=24$ bits,  is 
$\mathbf{x}=\color{blue}1110~0000~1101~0001~\color{red}0010~\color{red}0111\color{black}.$
For simplicity we skip the last encoding step (Block IV) intended to protect the parity bits and assume  that deletions affect only the systematic bits.
\end{example}

The high level idea of the decoding algorithm is to: (i)~make an assumption on in which block the bit deletion has occurred (the guessing part); (ii)~chunk the bits accordingly, treat the affected block as erased, decode the erasure and check whether the obtained sequence is consistent with the parities (the checking part); (iii)~go over  all the possibilities. 



\begin{example}[Successful Decoding] \label{ex:1a}
Suppose that the $14^{th}$ bit of $\mathbf{x}$ gets deleted. The  decoder receives the following $23$ bit  string $\mathbf{y}=\color{blue}111000001101001\color{red}00100111\color{black}.$ The decoder goes through all the possible $k/\log k=4$ cases, where in each case $i$, $i=1,\ldots,4$, the deletion is assumed to have occurred in block $i$ and $\mathbf{y}$ is chunked accordingly. Given this assumption, symbol $i$ is considered erased and erasure decoding is applied over $GF(16)$ to recover this symbol. Furthermore, given two parities, each symbol $i$ can be recovered in two different ways. Without loss of generality, we assume that the first parity $p_1$, $p_1=\alpha$, is the parity used for decoding the erasure. The decoded $q-$ary string in case $i$ is denoted by $\mathbf{Y_i}\in GF(16)^4$, and its binary representation is denoted by $\mathbf{y_i}\in GF(2)^{16}$. The four cases are shown below: \\
{\em \underline{Case 1}:} The deletion is assumed to have occurred in block 1, so $\mathbf{y}$ is chunked as follows and the erasure is denoted by  $\mathcal{E}$,
\begin{equation*}
\color{blue} \underbrace{1~1~1}_{\mathcal{E}}~\underbrace{0~0~0~0}_{0}~\underbrace{0~1~1~0}_{\alpha^{5}}~\underbrace{1~0~0~1}_{\alpha^{14}}~\color{red}\underbrace{0~0~1~0}_{\alpha}~\underbrace{0~1~1~1}_{\alpha^{10}}\color{black}.
\end{equation*}
Applying erasure decoding over $GF(16)$, the recovered value of symbol 1 is $\alpha^{13}$. Hence, the decoded $q-$ary string is $\mathbf{Y_1}=(\color{blue}\alpha^{13},0,\alpha^{5},\alpha^{14}\color{black})$. Its equivalent in binary is $\mathbf{y_1} = \color{blue}1101~0000~0110~1001\color{black}.$
Now, to check  our assumption, we test whether $\mathbf{Y_1}$ is consistent with the second parity $p_2=\alpha^{10}$. However, the computed parity is $\left(\color{blue}\alpha^{13},0,\alpha^{5},\alpha^{14}\color{black}\right)\left(1,\alpha,\alpha^{2},\alpha^{3}\right)^{T}=\alpha \neq \color{red} \alpha^{10}$.
This shows that $\mathbf{Y_1}$ does not satisfy the second parity. We deduce that our assumption on the deletion location  is wrong. Throughout the paper we refer to such cases as {\em impossible} cases.\\
{\em \underline{Case 2}:}  The deletion is assumed to have occurred in block 2, so the sequence is chunked as follows
\begin{equation*}
\color{blue} \underbrace{1~1~1~0}_{\alpha^{11}}~\underbrace{0~0~0}_{\mathcal{E}}~\underbrace{0~1~1~0}_{\alpha^{5}}~\underbrace{1~0~0~1}_{\alpha^{14}}~\color{red}\underbrace{0~0~1~0}_{\alpha}~\underbrace{0~1~1~1}_{\alpha^{10}}\color{black}.
\end{equation*}
Applying erasure decoding, the recovered value of symbol 2 is $\alpha^4$. Now, before checking whether the decoded string is consistent with the second parity $p_2$, one can notice that the binary representation of the decoded erasure $(0011)$ is not a supersequence of the sub-block $(000)$. So, without checking $p_2$, we can deduce that this case is {\em impossible}. 
\begin{definition} \label{def:1} We restrict this definition to the case of $\delta=1$ deletion with two MDS parity symbols in $GF(q)$. A case $i$, $i=1,2,\ldots,k/\log k$, is said to be {\em possible} if it satisfies the two criteria below simultaneously. \\
Criterion 1: The $q-$ary string decoded based on the first parity in case $i$, denoted by $\mathbf{Y_i}$, satisfies the second parity. \\
Criterion 2: The binary representation of the decoded erasure is a supersequence of its corresponding sub-block. \\
If any of the two criteria is not satisfied, the case is said to be {\em impossible}.
\end{definition}
\noindent The two criteria mentioned above are both necessary. For instance, in this example, case 2 does not satisfy Criterion 2 but it is easy to verify that it satisfies Criterion 1. Furthermore, case 1 satisfies Criterion 1 but does not satisfy Criterion 2. A case is said to be {\em possible} if it satisfies both criteria simultaneously. \\
\noindent {\em \underline{Case 3}:} By following the same steps as cases 1 and 2, it is easy to verify that both criteria are not satisfied in this case, i.e., case 3 is also {\em impossible}. \\
\noindent {\em \underline{Case 4}:} The deletion is assumed to have occurred in block 4, so the sequence is chunked as follows
\begin{equation*}
\color{blue} \underbrace{1~1~1~0}_{\alpha^{11}}~\underbrace{0~0~0~0}_{0}~\underbrace{1~1~0~1}_{\alpha^{13}}~\underbrace{0~0~1}_{\mathcal{E}}~\color{red}\underbrace{0~0~1~0}_{\alpha}~\underbrace{0~1~1~1}_{\alpha^{10}}\color{black}.
\end{equation*}
In this case, the decoded string is $\mathbf{y_4}=1110000011010001$. This case satisfies both criteria and is indeed {\em possible}. \\
 After going through all the cases, case 4 stands alone as the only {\em possible} case. So the decoder declares successful decoding and outputs $\mathbf{y_4}$ ($\mathbf{y_4}=\mathbf{u}$). 
\end{example}

\noindent The next example considers another message $\mathbf{u}$ and shows how the proposed decoding scheme can lead to a decoding failure. The importance of Theorem~\ref{thm:1} is that it shows that the probability of a decoding failure vanishes a $k$ goes to infinity.

\begin{example}[Decoding failure]
\label{ex:2}
Let  $\mathbf{u}=1101000010000101\color{black}.$
Following the same encoding steps as before, the $q-$ary codeword  is given by $\mathbf{X}=(\color{blue}\alpha^{13},0,\alpha^{3},\alpha^{8},\color{red}0,\alpha^{8}\color{black})$.  Suppose that the $14^{th}$ bit of the binary codeword $\mathbf{x}$ gets deleted. The decoder receives $\mathbf{y}=\color{blue} 110100001000001\color{red}00000101 \color{black}.$ The decoding is carried out as explained in Example~\ref{ex:1a}. The $q-$ary strings decoded in cases 1 and 4 are given by $\mathbf{Y_1}=(\color{blue}\alpha^{13},\alpha^{3},\alpha^{2},1\color{black})$ and $\mathbf{Y_4}=(\color{blue}\alpha^{13},0,\alpha^{3},\alpha^{8}\color{black})$, respectively. It is easy to verify that both cases 1 and 4 are {\em possible} cases. The decoder here cannot know which of the two cases is the correct one, so it declares a decoding failure. 
\end{example}
\section{General Decoding of GC codes}
\label{sec:5}
The encoding and decoding steps for $\delta>1$ deletions are a direct generalization of the steps for $\delta=1$ described previously. WLOG, we assume that exactly $\delta$ deletions have occurred. Then, the length of the string received by the decoder is  $n-\delta$ bits. Now, we explain in details the decoding steps. \\
\noindent $1)$ Decoding the parity symbols of Block II (Fig.~\ref{fig:1}): these parities are protected by a $(\delta+1)$ repetition code and therefore can be always recovered correctly by the decoder. Therefore, for the remaining steps we will  assume WLOG that all the $\delta$ deletions have occurred in the systematic bits.\\
\noindent $2)$ The guessing part: the number of possible ways to distribute the $\delta$ deletions among the $k/\log k$ blocks is $t=\binom{k/\log k+\delta-1}{\delta}.$ We index these possibilities by $i,i=1,\ldots,t,$ and refer to each possibility by case $i$. \\
\noindent The decoder goes through all the $t$ cases (guesses). \\
\noindent $3)$ The checking part: for each case $i$, $i=1,\ldots,t$, the decoder (i) chunks the sequence according to the corresponding assumption; (ii) considers the affected blocks erased and maps the remaining blocks to their corresponding symbols in $GF(q)$; (iii) decodes the erasures using the first $\delta$ parity symbols; (iv)~checks whether the case is {\em possible} or not based on the criteria described below. 
\begin{definition} \label{def:4} For $\delta$ deletions, a case $i$, $i=1,\ldots,t$, is said to be {\em possible} if it satisfies the following two criteria simultaneously. Criterion 1: the decoded $q-$ary string in case $i$ satisfies the last $c-\delta$ parities simultaneously. Criterion 2: the binary representations of all the decoded erasures are supersequences of their corresponding sub-blocks (for a given decoded erasure of length $\log k$ bits, the complexity of this is $\mathcal{O}(\log^2 k)$  using the Wagner-Fischer algorithm). 
\end{definition}
\noindent $4)$ After going through all the cases, the decoder declares successful decoding if (i) only one {\em possible} case exists; or (ii)~multiple {\em possible} cases exist but all lead to the same decoded string. Otherwise, a decoding failure is declared.
\vspace{-0.1cm}
\begin{remark}[Trade-offs]
\label{rem1}
GC codes enable two trade-offs. (1)~Decoding complexity and redundancy trade-off: We chose to chunk the message into blocks of $\log k$ bits in order to achieve an asymptotically optimal redundancy given by Levenshtein's bound. If the message is chunked into blocks of length $\ell$ bits, the redundancy becomes $c(\delta+1)\ell$ and the number of cases becomes $t=\binom{k/\ell+\delta-1}{\delta}$. The number of cases is the dominant factor in the decoding complexity. Therefore, by increasing $\ell$ the decoding complexity can be decreased while increasing the redundancy. Note that the probability of failure would still go to zero if $\ell=\Omega(\log k)$. (2)~Probability of failure and redundancy trade-off: The choice of the constant $c$ presents a trade-off between the redundancy and the probability of decoding failure. In fact, for a fixed $k$, by increasing $c$ the redundancy $c(\delta+1)\log k$ increases linearly while the probability of decoding failure $Pr(F)=\mathcal{O}\left(\frac{k^{2\delta-c}}{\log^{\delta}k}\right)$ decreases exponentially. Note that the order of complexity of the scheme is not affected by the choice of $c$.
\end{remark}
\vspace{-0.3cm}
\section{Proof of Theorem 1}
\label{sec:6}
In this section, we prove the upper bound on the probability of decoding failure $Pr(F)$ in Theorem~\ref{thm:1} for $\delta=1$ deletion. The complete and general proof follows similar steps and can be found in the extended version of this paper \cite{Ex}.  
The probability of decoding failure for $\delta=1$ is computed over all possible $k-$bit messages and all possible single deletions. Recall that the bits of the message $\mathbf{u}$ are i.i.d. Bernoulli$(1/2)$ and the position of the deletion is uniformly random. The message $\mathbf{u}$ is encoded as shown in Fig.~\ref{fig:1}. For $\delta=1$, the decoder goes through a total of $k/\log k$ cases, where in a case $i$ it decodes by assuming that block $i$ is affected by the deletion. Let $\bm{\mathcal{Y}_i}$ be the random variable representing the $q-$ary string decoded in case $i$, $i=1,2,\ldots,k/\log k$, in step 3 of the decoding scheme. Let $\mathbf{Y} \in GF(q)^{k/\log k}$ be a realization of the random variable $\bm{\mathcal{Y}_i}$.
We denote by $\mathcal{P}_r \in GF(q), r=1,2,\ldots,c,$ the random variable representing the $r^{th}$ MDS parity symbol (Block II, Fig.~\ref{fig:1}). Also, let $\mathbf{G_r} \in GF(q)^{k/\log k}$ be the MDS encoding vector responsible for generating $\mathcal{P}_r$. Consider $c>\delta$ arbitrary MDS parities $p_1,\ldots,p_c$, for which we define the following sets. For $r=1,\ldots,c,$
\begin{align*}
\mathrm{A_r} &\triangleq \{ \mathbf{Y} \in GF(q)^{k/\log k} |~\mathbf{G_r^TY}=p_r\}, \\
\mathrm{A} &\triangleq \mathrm{A_1} \cap \mathrm{A_2} \cap \ldots \cap \mathrm{A_c}.
\end{align*}
$\mathrm{A_r}$ and $\mathrm{A}$ are affine subspaces of dimensions $k/\log k -1$ and $k/\log k-c$, respectively. Therefore,
\begin{equation}
\label{e:A}
\left \lvert \mathrm{A_r} \right \rvert = q^{\frac{k}{\log k}-1} \text{ and} ~ \left \lvert \mathrm{A} \right \rvert = q^{\frac{k}{\log k}-c}. 
\end{equation}
Recall that the correct values of the MDS parities are recovered at the decoder, and that for $\delta=1$, $\bm{\mathcal{Y}_i}$ is decoded based on the first parity. Hence, for a fixed MDS parity $p_1$, and for $\delta=1$ deletion, $\bm{\mathcal{Y}_i}$ takes values in $\mathrm{A_1}$. Note that $\bm{\mathcal{Y}_i}$ is not necessarily uniformly distributed over $\mathrm{A_1}$. The crux of the proof relies on the next claim and its generalization. The claim gives an upper bound on the probability mass function of  $\bm{\mathcal{Y}_i}$ for $\delta=1$ deletion. Its proof can be found in \cite{Ex}.

\begin{claim} 
\label{claim:1}
For any case $i$, $i=1,2,\ldots,k/\log k$, $Pr\left(\bm{\mathcal{Y}_i}=\mathbf{Y} | \mathcal{P}_1=p_1 \right) \leq \frac{2}{q^{\frac{k}{\log k}-1}}$.

\end{claim}
\noindent  Claim~\ref{claim:1} can be interpreted as that at most $2$ different input messages can generate the same decoded string $\bm{\mathcal{Y}_i} \in \mathrm{A_1}$. Next, we use this claim to show that for $\delta=1$,
\begin{equation}
Pr(F) < \frac{2}{k^{c-2}~\log k}. \label{d:1}
\end{equation}
\noindent In the general decoding scheme, we mentioned two criteria which determine whether a case is {\em possible} or not (Definition~\ref{def:4}). Here, we upper bound $Pr(F)$ by taking into account Criterion~$1$ only. Based on Criterion~$1$, if a case $i$ is possible, then $\bm{\mathcal{Y}_i}$ satisfies all the $c$ MDS parities simultaneously, i.e., $\bm{\mathcal{Y}_i} \in \mathrm{A}$. Without loss of generality, we assume case~$1$ is the correct case, i.e., the deletion occurred in block~$1$. A decoding failure is declared if there exists a {\em possible} case $j$, $j=2,\ldots,k/\log k$, that leads to a decoded string different than that of case~$1$. Namely, $\bm{\mathcal{Y}_j} \in \mathrm{A}$ and $\bm{\mathcal{Y}_j} \neq \bm{\mathcal{Y}_1}$. Therefore,
\begin{align}
Pr\left(F | \mathcal{P}_1=p_1 \right) &\leq Pr\left(\bigcup_{j=2}^{k/\log k}\{\bm{\mathcal{Y}_j}\in \mathrm{A},\bm{\mathcal{Y}_j} \neq \bm{\mathcal{Y}_1}\} \biggr\rvert \mathcal{P}_1=p_1 \right) \\
&\leq \sum_{j=2}^{k/\log k} Pr\left(\bm{\mathcal{Y}_j}\in \mathrm{A},\bm{\mathcal{Y}_j} \neq \bm{\mathcal{Y}_1} | \mathcal{P}_1=p_1 \right) \label{e:2} \\
&\leq \sum_{j=2}^{k/\log k} Pr\left(\bm{\mathcal{Y}_j}\in \mathrm{A} | \mathcal{P}_1=p_1 \right) \label{e:4}\\
&= \sum_{j=2}^{k/\log k} \sum_{\mathbf{Y}\in \mathrm{A}} Pr\left(\bm{\mathcal{Y}_j}=\mathbf{Y} | \mathcal{P}_1=p_1 \right) \\
&\leq \sum_{j=2}^{k/\log k}  \left \lvert \mathrm{A} \right \rvert \frac{2}{q^{\frac{k}{\log k}-1}} \label{e:6} \\
&< \frac{2}{k^{c-2}~\log k} \label{e:10}.
\end{align}
\noindent \eqref{e:2} follows from applying the union bound. \eqref{e:4} follows from the fact that $Pr\left(\bm{\mathcal{Y}_j} \neq \bm{\mathcal{Y}_1}|\bm{\mathcal{Y}_j}\in \mathrm{A}, \mathcal{P}_1=p_1 \right)\leq 1$. \eqref{e:6} follows from Claim~\ref{claim:1}. \eqref{e:10} follows from \eqref{e:A} and the fact that $q=k$ in the coding scheme. The proof of \eqref{d:1} is completed using \eqref{e:10} and averaging over all values of $p_1$.
\vspace{-0.1cm}
\section{Simulation Results}
\label{sec:7}

We simulated the decoding  of GC codes and compared the obtained probability of decoding failure  to the  upper bound in Theorem~\ref{thm:1}.
 We tested the code for messages of length $k~=~256, 512$ and $1024$ bits, and for $\delta=2, 3$ and $4$ deletions. 

\begin{table}[h]
\centering
\setlength\extrarowheight{1.2pt}
 \begin{tabular}{|c|c|c|c|c|c|c|c|c|}
\hline
\multirow{2}{*}{Config.} & \multicolumn{6}{c|}{$\delta$} \\ \cline{2-7}
& \multicolumn{2}{c|}{$2$}  & \multicolumn{2}{c|}{$3$} &  \multicolumn{2}{c|}{$4$} \\ \hline
$k$ & $R$ & $Pr(F)$ & $R$ & $Pr(F)$ &  $R$ & $Pr(F)$ \\ \hline
256 & $0.78$ & $1.3e^{-3}$ & $0.67$ & $4.0e^{-4}$ & $0.56$ &$0$ \\ \hline
512 & $0.86$ & $3.0e^{-4}$ & $0.78$ &$0$ & $0.69$ &$0$  \\ \hline
1024 & $0.92$ & $2.0e^{-4}$ & $0.86$ &$0$ & $0.80$ &$0$  \\ \hline
\end{tabular}
\captionsetup{font=footnotesize}
\caption{\footnotesize{The table shows the code rate $R=k/n$ and the probability of decoding failure $Pr(F)$  of GC codes for different message lengths $k$ and different number of deletions $\delta$. The results of $Pr(F)$ are averaged over $10000$ runs of simulations. 
\vspace{-0.6cm}
}}
\label{t}
\end{table}
\noindent 
To guarantee an asymptotically vanishing probability of decoding failure, the upper bound in Theorem~\ref{thm:1} requires that $c>2\delta$. Therefore, we make a distinction between two regimes, \mbox{(i) $\delta<c<2\delta:$}~Here, the theoretical upper bound is trivial. Table~\ref{t} gives the results for $c=\delta+1$ with the highest probability of decoding failure observed in our simulations  being of the order of $10^{-3}$. 
This indicates that GC codes can decode correctly with high probability in this regime, although not reflected in the upper bound;  \mbox{(ii) $c>2\delta:$}  The upper bound is of the order of $10^{-5}$ for $k=1024,\delta=2$, and $c=2\delta +1$. In the simulations no decoding failure was detected within $10000$ runs for $\delta+2 \leq c \leq 2\delta+1$. 
In general, the simulations show that GC codes  perform better than what the upper bound indicates. This is due to the fact that the effect of Criterion 2 (Definition~\ref{def:4}) is not taken into account when deriving the upper bound in Theorem~\ref{thm:1}. These simulations were performed on a personal computer. The average decoding time is in the order of milliseconds  for $(k=1024,\delta=2)$, order of seconds for $(k=1024,\delta=3)$, and order of minutes for $(k=1024,\delta=4)$. Going beyond these values of $k$ and $\delta$ will largely increase the running time due to the number of cases to be tested by the decoder. However, for the file synchronization application in which we are interested (see next section) the values $k$ and $\delta$ are relatively small and decoding can be practical. 

\section{Application to File Synchronization}
\label{sec:8}
In this section, we describe how our codes can be used to construct interactive protocols for file synchronization. We consider the model where two nodes (servers) have copies of the same file but one is obtained from the other by deleting $d$ bits. These nodes communicate interactively over a noiseless link to synchronize the file affected by deletions. Some of the most recent work on synchronization can be found in \cite{V15,Y14}. In this section, we modify the synchronization algorithm by Venkataramanan {\em et al.} \cite{V15}, and study the improvement that can be achieved by including our code as a black box inside the algorithm. The key idea in \cite{V15} is to use {\em center bits} to divide a large string, affected by $d$ deletions, into shorter segments, such that each segment is affected by only one deletion. Then, use VT codes to correct these segments. Now, consider a similar algorithm where the large string is divided such that the shorter segments are affected by $\delta$ $(1<\delta\ll d)$ or fewer deletions. Then, use the GC code to correct the segments affected by more than one deletion.
We set $c=\delta+1$, and if the decoding fails for a certain segment, we send one extra MDS parity at a time within the next communication round until the decoding is successful. By implementing this algorithm, the gain we get is two folds: (i)~reduction in the number of communication rounds; (ii)~reduction in the total communication cost. We performed simulations for $\delta=2$ on files of size 1 Mb, for different numbers of deletions $d$. The results are illustrated in Table~\ref{ts}. We refer to the original scheme in \cite{V15} by {\em Sync-VT}, and to the modified version by {\em Sync-GC}. The savings for $\delta=2$ are roughly $43\%$ to $73\%$ in number of rounds, and $5\%$ to $14\%$ in communication cost. 
\begin{table}[h]
\centering
\setlength\extrarowheight{1.2pt}
 \begin{tabular}{|c|c|c|c|c|}
\hline
 & \multicolumn{2}{c|}{Number of rounds} & \multicolumn{2}{c|}{Total communication cost}\\ \hline
$d$ & Sync-VT & Sync-GC & Sync-VT & Sync-GC  \\ \hline
100 & $14.52$ & $10.15$ & $5145.29$ & $4900.88$ \\ \hline
150 & $16.45$ & $10.48$ & $7735.32$ & $7199.20$  \\ \hline
200 & $17.97$ & $10.88$ & $10240.60$ & $9332.68$  \\ \hline
250 & $18.93$ & $11.33$ & $12785.20$ & $11415.90$ \\ \hline
300 & $20.29$ & $11.70$ & $15318.20$ & $13397.80$ \\ \hline
\end{tabular}
\captionsetup{font=footnotesize}
\caption{ Results are averaged over $1000$ runs. In each run, a string of size 1~Mb is chosen uniformly at random, and the file to be synchronized is obtained by deleting $d$ bits from it uniformly at random. The communication cost is expressed in bits. The number of {\em center bits} used is 25.}
\label{ts}
\end{table}
\vspace{-0.38cm}

\section{Acknowledgments}
\noindent  The authors would like to thank Kannan Ramchandran for valuable discussions related to an earlier draft of this work.

\bibliographystyle{ieeetr}

\bibliography{Refs}

\end{document}